\documentclass[aps,pre,showpacs%
        ,amsmath%
        ,superscriptaddress%
        ,reprint%
        ,nofootinbib
        ,preprintnumbers
        ]{revtex4-1}

\usepackage[pdftex]{graphicx}
\usepackage{amsmath,amssymb}

\newcommand{\paulivec}{\bvec{\sigma}}
\newcommand{\bbset}[1]{\mathbb{#1}}
\newcommand{\Real}{\bbset{R}}
\newcommand{\set}[1]{\left\{ #1 \right\}}
 \newcommand{\ket}[1]{|{}#1{}\rangle}
 \newcommand{\ketbra}[2]{|{}#1{}\rangle\langle{}#2{}|}
\newcommand{\bvec}[1]{\boldsymbol{#1}} 

\newcommand{\MP}{\mathcal{P}} 
\newcommand{\MM}{\mathcal{M}} 

\newcommand{\RPP}{\Real{}P^2} 

\begin{document}

\title{A topological formulation for exotic quantum 
  holonomy}

\author{Atushi Tanaka}
\homepage[]{\tt http://researchmap.jp/tanaka-atushi/}
\affiliation{Department of Physics, Tokyo Metropolitan University, Hachioji, Tokyo 192-0397, Japan}

\author{Taksu Cheon}
\affiliation{Laboratory of Physics, Kochi University of Technology, Tosa Yamada, Kochi 782-8502, Japan}

\begin{abstract}
An adiabatic change of parameters along a closed path
may interchange the (quasi-)eigenenergies
and eigenspaces of a closed quantum system. Such discrepancies induced
by adiabatic cycles are refereed to as the exotic quantum holonomy,
which is an extension of the geometric phase.
``Small'' adiabatic cycles induce no change on eigenspaces, whereas
some ``large'' adiabatic cycles interchange eigenspaces. 
We explain the topological formulation for the eigenspace anholonomy,
where the homotopy equivalence precisely
distinguishes the larger cycles from smaller ones.
An application to two level systems is explained.
We also examine the cycles that involve the adiabatic evolution across an
exact crossing, and the diabatic evolution across an avoided
crossing. The latter is a nonadiabatic example of the exotic 
quantum holonomy.
\end{abstract}


\maketitle

\section{Introduction}

An adiabatic quasi-static cycle may induce a nontrivial change on
a closed quantum system. A well-known example
is the geometric phase factor~\cite{Berry-PRSLA-392-45,Shapere-GPP-1989}, 
which is also called as
the quantum holonomy because of its geometrical 
interpretation~\cite{Simon-PRL-51-2167,GPBook}.

Recently, it has been recognized that the geometric phase has exotic brothers, where (quasi-)eigenenergies and eigenspaces of stationary states exhibit nontrivial change as a result of an adiabatic cycle. Namely, eigenenergies and eigenspaces may be interchanged by adiabatic cycles. 

The exotic quantum holonomy has been found in various physical systems: a particle confined in a one-dimensional box with a generalized pointlike potential~\cite{Cheon-PLA-248-285}, a quantum map under a rank-$1$ perturbation~\cite{Tanaka-PRL-98-160407}, the Lieb-Liniger model\cite{Yonezawa-PRA-87-062113}, and a quantum graph~\cite{Cheon-ActaPolytechnica-53-410}. Other examples are reported in the references cited in Ref.~\cite{Tanaka-PLA-379-1693}.

In the following, we will briefly explain the topological formulation of the exotic quantum holonomy~\cite{Tanaka-PLA-379-1693}, which may be considered as a counterpart of the geometrical formulation for the geometric phase factor~\cite{Simon-PRL-51-2167,Aharonov-PRL-58-1593}. Using the homotopic classification, we discuss ``large'' cycles that exhibit exotic quantum holonomy in Hamiltonian systems that involve the exact level crossing and avoided crossing.

\section{Topological formulation}

The quantum holonomies including both the geometric phase and eigenspace anholonomy are discrepancies induced by an adiabatic cycle,
i.e., a closed path in an adiabatic parameter space $\MM$.
A lifting of the adiabatic cycle is helpful to characterize the quantum 
holonomies.

As for the geometric phase, we define a lift of the adiabatic path 
as the trajectory of the state 
vector, which satisfies the adiabatic time-dependent Schr\"odinger equation with
an adiabatic initial state.
We assume that the dynamical phase is excluded from the lift.

The lift of an adiabatic cycle $C$ induces a mapping, which
is denoted by $\phi_{C}$, from the initial adiabatic state 
to the final adiabatic state.
$\phi_{C}$ puts a
geometric phase factor to the initial state vector.
In terms of differential geometry, $\phi_C$ is
an element of the holonomy group.
This fact allows us to thoroughly investigate the geometric phase
with the help of differential geometry~\cite{Simon-PRL-51-2167}.

\begin{figure}[htbp]
  \begin{center}
  \includegraphics[width=4.2cm]{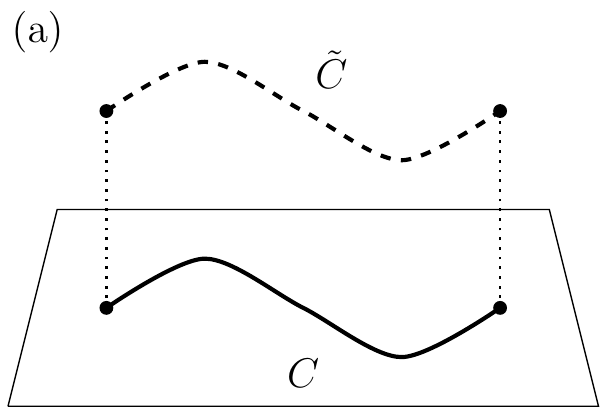}
  \\[2\baselineskip]
  \includegraphics[width=5.2cm]{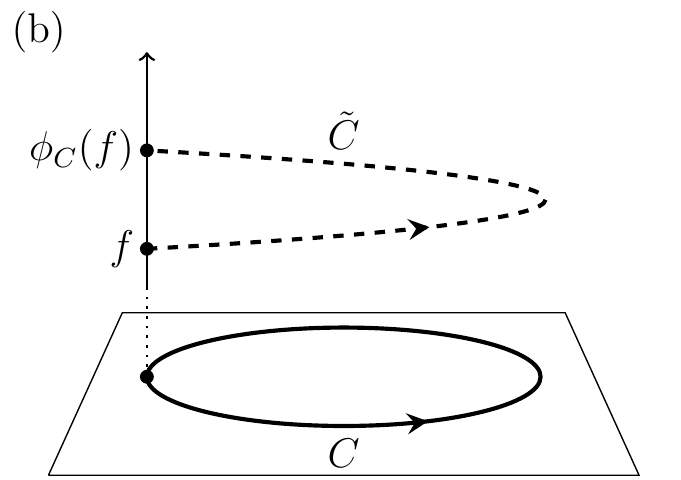}
  \end{center}
  \caption{Lifting structure behind quantum holonomies.
    (a) Lift $\tilde{C}$ (dashed line) of a path $C$ (bold line) in 
    the adiabatic parameter space $\MM$.
    A lift is the trajectory of an eigenobject under the adiabatic time
    evolution with an adiabatic initial condition.
    The dotted lines represent the projection $\pi$ from the lifted space
    to $\MM$. 
    As for the case of the geometric phase, the lift is 
    the trajectory of the state vector excluding the dynamical phase.
    On the other hand, the lift for the exotic quantum holonomy
    is the trajectory of $p$ (Eq.~\eqref{eq:def_p}).
    (b) Lift of the closed path (i.e., adiabatic cycle)
    for a given initial eigenobject~$f$.
    We denote the final point of the lift by $\phi_C(f)$.
    The mapping $\phi_C$ describes the discrepancy of the eigenobject 
    induced by the cycle $C$.
  }
\label{fig:lift}
\end{figure}

We carry over the concept of lifting to the exotic quantum holonomy.
Instead of the adiabatic state vector for the geometric phase, 
we employ an ordered set of eigenprojectors
\begin{equation}
  \label{eq:def_p}
  p \equiv (\hat{P}_1, \hat{P}_2, \ldots)
\end{equation}
where $\hat{P}_j$ is $j$-th eigenprojector under a given value of
the adiabatic parameter. We denote $p$-space by $\MP$.

We explain two kinds of the adiabatic parameter space $\MM$, which
locates at the bottom of the lifting structure. One is 
a $c$-number parameter space. On the other hand, 
it is useful to introduce a canonical 
``adiabatic parameter space'', whose point is a set of projectors
\begin{equation}
  b \equiv \{\hat{P}_1, \hat{P}_2, \ldots\},
\end{equation}
where the order of the projectors are disregarded. $b$-space is a counterpart
of the projective Hilbert space in the Aharonov-Anandan theory 
of the geometric phase~\cite{Aharonov-PRL-58-1593}.

We explain the lifting of an adiabatic cycle $C$ in $\MM$ to $\MP$, and
the corresponding $\phi_C$.
For a given adiabatic path, the adiabatic Schr\"odinger equation induces 
the adiabatic time evolution of $p$. 
The lifted path is the trajectory
of $p$, and naturally induces $\phi_{C}$. 
The mapping $\phi_{C}$ between the initial and final $p$ is
essentially a permutation of $\hat{P}_j$'s. 
For example, if $C$ does not induces the exotic quantum holonomy,
$\phi_{C}$ is equivalent with the identical permutation. 
On the other hand, $\phi_{C}$ may be a non-identical permutation,
which describes the interchange of the eigenspaces, induced by $C$.

Here our task is to characterize $\phi_{C}$ completely.
We utilize the fact that there is a covering map $\pi: \MP \to\MM$,
where $\pi$ is the projection that satisfies the axiom of the covering
projection (not shown here). 
We remark that the covering space
is a fiber bundle with a discrete 
structure group due to
the discreteness of $p$ for a given value
of the adiabatic parameter.

The covering map structure allows us to investigate the exotic quantum
holonomy
with the help of topology.
First, $\phi_{C}$ and $\phi_{C'}$ are same if $C$ and $C'$ are homotopic,
where we say that a cycle $C$ is homotopic to another cycle $C'$
if $C$ can be smoothly deformed to $C'$ with the initial and final points
kept unchanged.
Hence we may denote the mapping $\phi_{C}$ as $\phi_{[C]}$, where
$[C]$ is the class of paths that are homotopic to $C$.

Next, we need to enumerate all possible $[C]$'s in the adiabatic parameter
space $\MM$. This is equivalent to find 
$\pi_1(\MM)=\{[C]|\  \text{$C$ is a closed path in $\MM$}\}$,
which is called the first fundamental group of $\MM$.

Hence it suffices to enumerate $\phi_{[C]}$ for all $[C]\in\pi_1(M)$.
When $\MP$ is contractable to a point (i.e., there is no ``hole''),
there is a one-to-one correspondence between 
$\pi_1(\MM)$ and $\phi_{[C]}$, i.e.,
\begin{equation}
  \label{eq:phiVSpi}
  \{\phi_{[C]}\}_{[C]\in\pi_1(\MM)}\simeq\pi_1(\MM)
  .
\end{equation}
In other words, if $C$ is not homotopic to $C'$,
two permutations $\phi_{[C]}$ and $\phi_{[C']}$ are different.
The extension of Eq.~\eqref{eq:phiVSpi} to an arbitrary $\MP$ is
shown in Ref.~\cite{Tanaka-PLA-379-1693}.

\section{Analysis of two level systems}
We examine the exotic quantum holonomy in two level systems. 
For a while, we suppose the system has no spectral degeneracy. 
Hence the Hamiltonian $\hat{H}$ has the following spectral decomposition
\begin{equation}
  \label{eq:spectralDecomposition}
  \hat{H} = E_1 \hat{P}_1 + E_2 \hat{P}_2.
\end{equation}
where $E_1$ and $E_2$ are eigenvalues of $H$.
In two level systems, the eigenprojections $\hat{P}_1$ and $\hat{P}_2$
may be specified by
a normalized $3$-vector $\bvec{a}$, which is called as the Bloch vector, as
\begin{equation}
  \label{eq:eigenprojectors}
  \hat{P}_1 = \frac{1}{2}(1+\bvec{a}\cdot\hat{\paulivec})
  ,\quad
  \hat{P}_2 = \frac{1}{2}(1-\bvec{a}\cdot\hat{\paulivec})
  .
\end{equation}
Non-degenerate periodically driven systems can be examined in the same
manner once we replace $\hat{H}$ with a Floquet operator.

\begin{figure}[htbp]
  \begin{center}
    \begin{minipage}[t]{2.0cm}
      \vspace{0pt}
        \includegraphics[width=\textwidth]{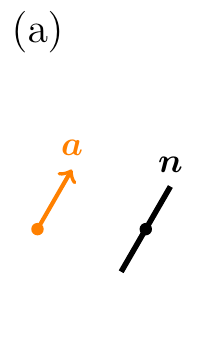}
        \vfill
    \end{minipage}
    \hspace{1em}
    \begin{minipage}[t]{5.5cm}
      \vspace{0pt}
        \includegraphics[width=\textwidth]{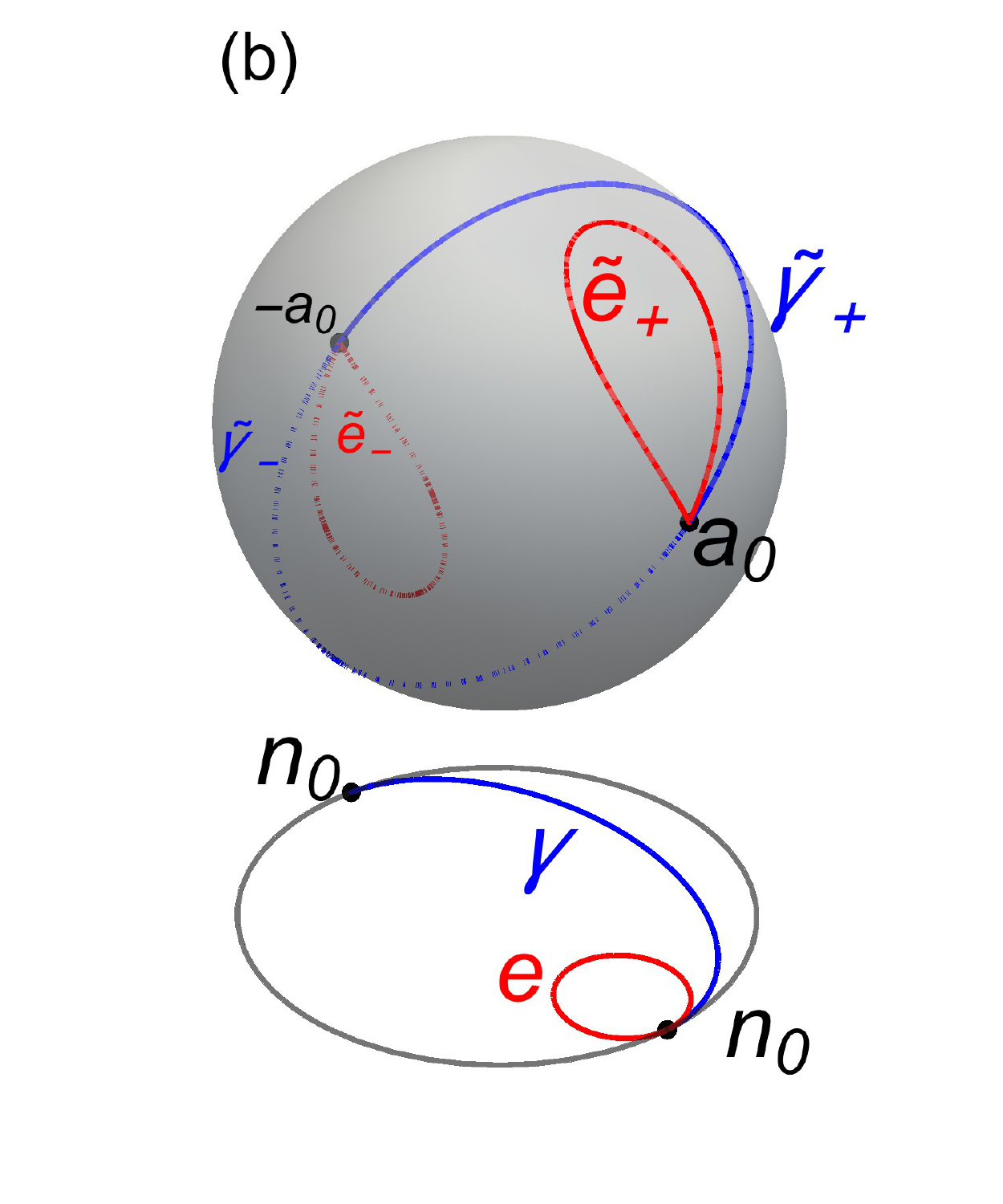}
    \end{minipage}
  \end{center}
  \caption{%
    (a) Bloch vector $\bvec{a}$ and director $\bvec{n}$ (schematic).
    (b)
    Two homotopically inequivalent cycles $e$ and $\gamma$ in $\RPP$ (bottom), 
    and their lifts to $S^2$ (upper).
    The initial point of the cycles is denoted as $\bvec{n}_0$.
    The lifts of the cycles with initial point $\pm\bvec{a}_0$
    are $\tilde{e}_{\pm}$ and $\tilde{\gamma}_{\pm}$.
    Since the former are closed in $S^2$, $\phi_{[e]}$ 
    correspond to the identical permutation of eigenspaces.
    On the other hand, the latter are open so that
    $\phi_{[\gamma]}$ corresponds to the cyclic permutation of
    the two eigenspaces, i.e., $\gamma$
    induces the exotic quantum holonomy.
   }
\label{fig:RP2andS2}
\end{figure}

We explain that $p$ and $b$ have a simple geometrical interpretation 
in two level systems (see, Fig.~\ref{fig:RP2andS2} (a)).
First, $p$ is equivalent to $\bvec{a}$, a point in a sphere $S^2$,
as is seen from Eq.~\eqref{eq:eigenprojectors}.
Second, $b$ is equivalent to the director (headless vector) 
$\bvec{n}$~\cite{Director},
which correspond to a point in the real projective plane $\RPP$.
This is because both $\bvec{a}$ and $-\bvec{a}$ correspond to
the same value of $b$, and the identification of antipodal point on
the sphere leads to $\RPP$.

Once we employ $\bvec{n}$ as the adiabatic parameter space, the topological
formulation provides an intuitive interpretation of the exotic quantum 
holonomy. The lifts of a ``small'' cycle in $\bvec{n}$-space are
closed in $\bvec{a}$-space, i.e., the sphere (Fig.~\ref{fig:RP2andS2} (b)).
On the other hand, the lifts of a ``large'' cycle in $\bvec{n}$-space can
be open in $\bvec{a}$-space. The exotic quantum holonomy reflects such 
a discrepancy between the trajectories of $\bvec{n}$ and $\bvec{a}$.

A complete way to distinguish the smaller and larger cycles is provided
by the homotopy equivalence.
In $\bvec{n}$-space, i.e., $\RPP$, there are 
only two classes of cycles, i.e.,
$\pi_1(\RPP)=\set{[e], [\gamma]}$,
where
$e$ is a ``small'' cycle that is homotopic to a point
(i.e., the zero-length closed path), and
a ``large'' cycle $\gamma$ is not homotopic to $e$  
(Fig.~\ref{fig:RP2andS2} (b)).

Now
Eq.~\eqref{eq:phiVSpi} is applicable to classify $\phi_{[C]}$ completely,
since $\bvec{a}$-space, i.e. $S^2$, has no hole.
Namely, $\phi_{[e]}$ and $\phi_{[\gamma]}$ correspond to different
permutations, the identical and the cyclic permutations of two items,
respectively (see, Fig.~\ref{fig:RP2andS2} (b)).

\section{Examples of nontrivial cycle $\gamma$}
We explain how we realize the adiabatic cycles $\gamma$, which 
induces the cyclic permutation of eigenspaces as well as 
(quasi-)eigenenergies in non-degenerate two level systems.

\begin{figure}[htbp]
  \centerline{\includegraphics[width=3.6cm]{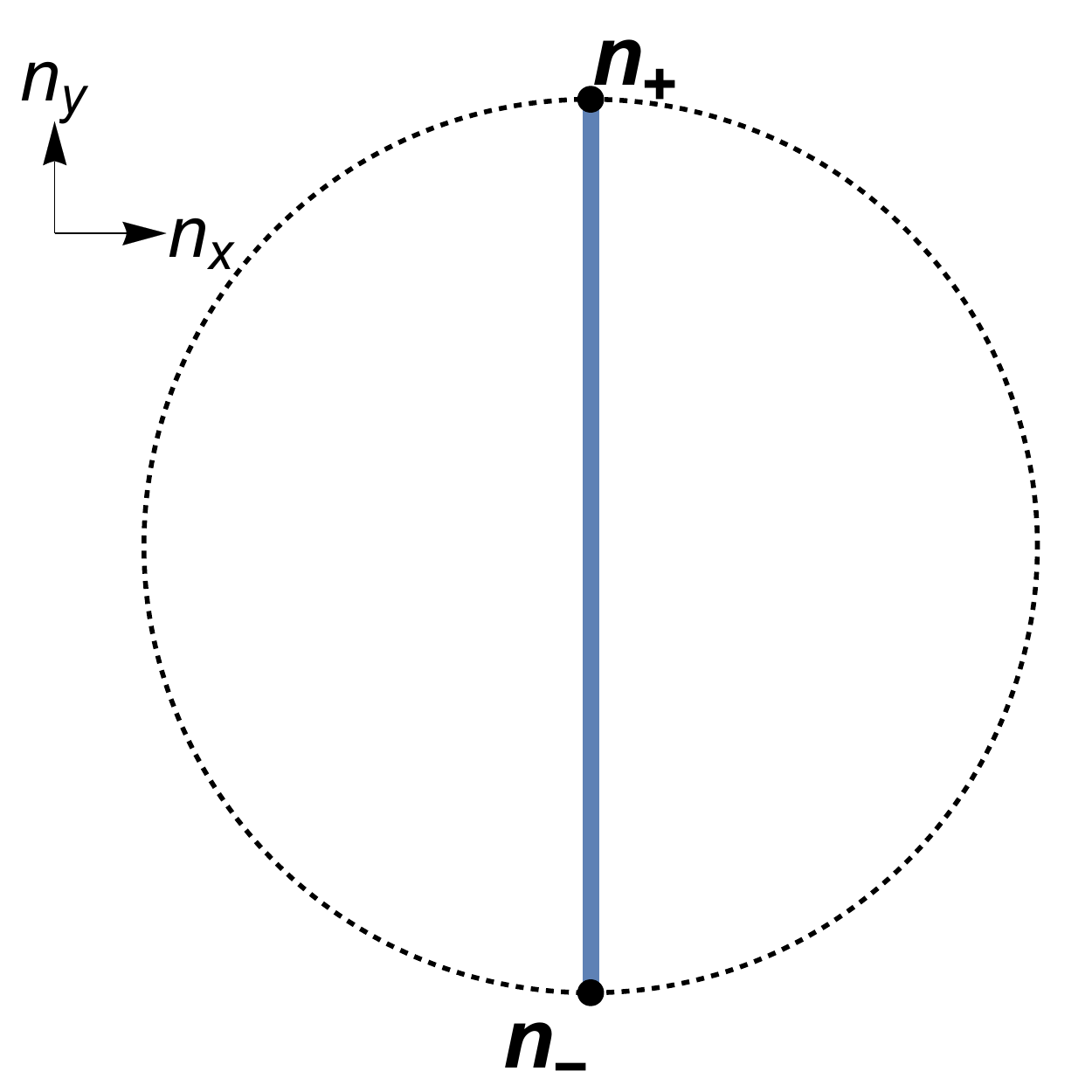}}
  \caption{Trajectories of the director $\bvec{n}$.
    We depict the trajectories in $n_x$-$n_y$ plane.
    The first example is the quantum map~\eqref{eq:defU}
    with $\hat{H}_0=\frac{\pi}{2}\hat{\sigma}_y$ 
    and $\ketbra{v}{v} = (1+\hat{\sigma}_x)/2$,
    which implies 
    $\bvec{a}=\bvec{e}_y\cos(\lambda/2)-\bvec{e}_z\sin(\lambda/2)$,
    whose director moves along $n_y$-axis.
    Note that the initial and end points $\bvec{n}_{\pm}$ are
    same in $\RPP$.
    Hence the trajectory is closed, and homotopic to $\gamma$.
    The Bloch vector of the second example~\eqref{eq:Hdegenerate} 
    is shown in Eq.~\eqref{eq:2ndBloch}. The trajectory of 
    the director coincides with the one of the first example,
    and agrees with the one for the third nonadiabatic example.
  }
  \label{fig:trajectory_of_n}
\end{figure}

The first example is a family of quantum map under a rank-1 perturbation,
which is described by a periodically kicked Hamiltonian
$
\hat{H}(\lambda,t) 
= \hat{H}_0 + \lambda\ketbra{v}{v}\sum_{n=-\infty}^{\infty}\delta(t-n)
$,
where $\ket{v}$ is a normalized vector. We introduce a Floquet operator,
which describes the time evolution during a unit time interval,
\begin{equation}
  \label{eq:defU}
  \hat{U}(\lambda) = e^{-i\hat{H}_0}e^{-i\lambda\ketbra{v}{v}}
  .
\end{equation}
The stationary state of this system for a given value of $\lambda$
is described by an eigenvector
of $\hat{U}(\lambda)$, and the time evolution induced by an 
adiabatic variation of $\lambda$ is essentially governed by
the parametric evolution of eigenvectors of 
$\hat{U}(\lambda)$~\cite{DiscreteAdiabatic}.
Since $\hat{U}(\lambda)$ has a period $2\pi$ as a function of $\lambda$,
the trajectory of $\bvec{n}$ induced by $\hat{U}(\lambda)$ for
$0\le \lambda\le 2\pi$ makes a closed path in $\RPP$.
Nevertheless, the lift of the closed path to $S^2$ is generically open
to exhibit the non-identical permutation of eigenprojectors,
as these closed paths are homotopic to $\gamma$ 
(Fig.~\ref{fig:trajectory_of_n}).
See, Refs.\cite{Tanaka-PRL-98-160407,Tanaka-PLA-379-1693} for details.

\begin{figure}[htbp]
  \centerline{\includegraphics[width=6.0cm]{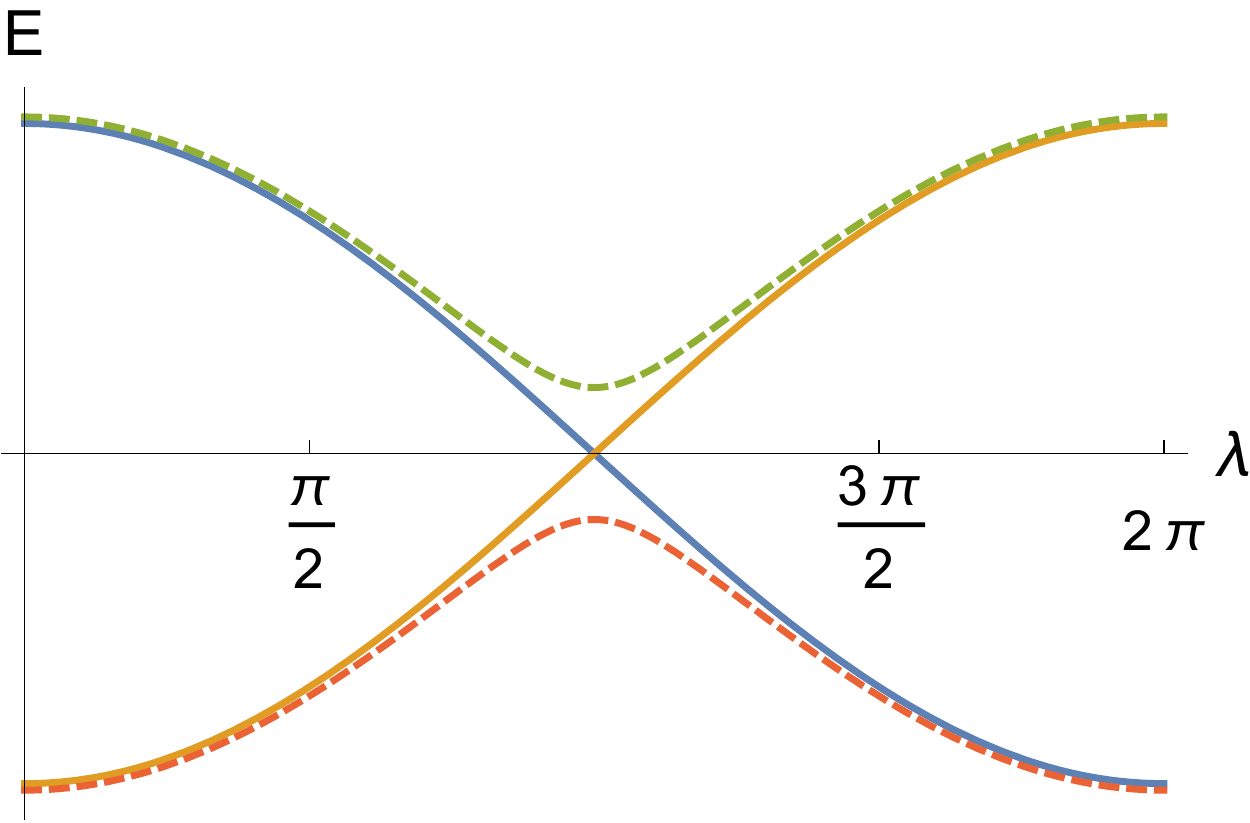}}
  \caption{Eigenenergies of the level crossing 
    Hamiltonian $\hat{H}(\lambda)$ (Eq.~\eqref{eq:Hdegenerate}) (thick lines)
    and the perturbed Hamiltonian $\hat{H}_{\epsilon}(\lambda)$
    (Eq. \eqref{eq:Hperturbed}) (dashed lines).
    Since the eigenenergies of $\hat{H}(\lambda)$ crosses at $\lambda=\pi$,
    the adiabatic cycle $0\le \lambda \le 2\pi$ interchanges
    the eigenspaces at $\lambda=0$.
    The adiabatic cycle for $\hat{H}_{\epsilon}(\lambda)$ do not induce the
    exotic quantum holonomy, due to the presence of the avoided crossing
    around $\lambda=\pi$.
    However, if $\lambda$ is moved quickly around $\lambda=\pi$, 
    the system follows the diabatic evolution. Such a diabatic cycle
    induces the exotic quantum holonomy
    along the nonadiabatic time evolution
    ~\cite{Cheon-PLA-374-144}.
  }
  \label{fig:levels}
\end{figure}

Second, we examine the adiabatic cycles that involve a level 
crossing~\cite{Cheon-PLA-374-144} using the following Hamiltonian
\begin{equation}
  \label{eq:Hdegenerate}
  \hat{H}(\lambda) 
  \equiv \frac{1}{4}\left[(1+\cos\lambda)\hat{\sigma}_y+(\sin\lambda)\hat{\sigma}_z\right]
  .
\end{equation}
which is periodic in $\lambda$ with a period $2\pi$, and
degenerates at $\lambda=\pi$. 
We introduce a Bloch vector
\begin{equation}
  \label{eq:2ndBloch}
  \bvec{a}
  \equiv \cos(\lambda/2)\bvec{e}_y+\sin(\lambda/2)\bvec{e}_z
\end{equation}
for Eq.~\eqref{eq:Hdegenerate}, which leads to the spectrum 
decomposition~\eqref{eq:spectralDecomposition} with
the eigenprojectors~\eqref{eq:eigenprojectors},
$E_1 = \frac{1}{2}\cos(\lambda/2)$ and 
$E_2 = -\frac{1}{2}\cos(\lambda/2)$ (Fig.~\ref{fig:levels}).
We note that the Bloch vector $\bvec{a}$ smoothly depends on $\lambda$
even in the vicinity of the crossing point $\lambda=\pi$. 
Hence the adiabatic time evolution follows
the corresponding parametric evolution of eigenprojectors
~\eqref{eq:eigenprojectors}~\cite{Kato-JPSJ-5-435}.
Although $\hat{H}(\lambda)$ is periodic in $\lambda$ with the period $2\pi$, 
the eigenenergies as well as the eigenprojectors are not,
which implies the presence of the exotic quantum holonomy.
Hence the trajectory of the director for $0\le \lambda\le2\pi$ is 
closed, and is homotopic to $\gamma$ (see, Fig.~\ref{fig:trajectory_of_n}).

Third, we examine the second case under a generic small perturbation,
which generically breaks the spectral degeneracy. For example, 
we consider
\begin{equation}
  \label{eq:Hperturbed}
  \hat{H}_{\epsilon}(\lambda) 
  \equiv \hat{H}(\lambda) + \frac{1}{2}\epsilon\sigma_x
  .
\end{equation}
The level crossing point becomes an avoided crossing point due to 
the perturbation (Fig.~\ref{fig:levels}).
Hence the exotic quantum holonomy do not occur in the adiabatic limit.
However, the use of the diabatic process across
the avoided crossing recovers the exotic quantum 
holonomy~\cite{Cheon-PLA-374-144}.
Along the closed path $0\le\lambda\le2\pi$ involving the diabatic process,
the trajectory induced by the time evolution of the state projector, 
which is initially an eigenprojector of $\hat{H}(0)$, mimics 
the parametric evolution of the corresponding eigenprojector
of the unperturbed system $\hat{H}(\lambda)$.
In this sense, the diabatic cycle plays the role of nontrivial cycle $\gamma$.
We remark that this is a nonadiabatic example of 
the exotic quantum holonomy.

\section{Summary}
We briefly explained 
the topological formulation of the exotic quantum 
holonomy with an emphasis to 
its geometrical character in two level systems.
An application to Hamiltonian systems with an exact or avoided crossing
is explained.

\section*{Acknowledgements}
This research was supported by the Japan Ministry of Education, Culture, Sports, Science and Technology under the Grant number 15K05216.



%

\end{document}